\documentclass[]{article}
\usepackage[left=2.5cm, right=2.5cm, top=1.785cm, bottom=2.0cm]{geometry} 
\usepackage{graphicx}  
\usepackage[parfill]{parskip} 
\usepackage{endnotes} 
\usepackage{setspace} 
\usepackage{filecontents}
\usepackage{amsmath}
\usepackage{amssymb}
\usepackage{palatino}
\usepackage{gensymb}
\usepackage{authblk}
\usepackage{wrapfig}
\usepackage{float}

\title{Interparticle hydrogen bonding can elicit shear jamming in dense suspensions}
\author[a,b]{Nicole James}
\author[a,c]{Endao Han}
\author[a]{Justin Jureller}
\author[a,c]{Heinrich Jaeger}
\affil[a]{James Franck Institute, The University of Chicago, 929 E 57th St, Chicago, USA}
\affil[b]{Department of Chemistry, The University of Chicago, 5735 S. Ellis Ave, Chicago, USA}
\affil[c]{Department of Physics, The University of Chicago, 5720 S. Ellis Ave, Chicago, USA}
\date{May 26, 2017}

\begin{document}

\maketitle

\hrule
\section*{Abstract}
Dense suspensions of hard particles in a liquid can exhibit strikingly counter-intuitive behavior, such as discontinuous shear thickening (DST) \cite{Barnes1989,Seto2013,Fernandez2013,Cates-friction,Brown2014Rev,Mari2015,Royer2015, Mari2014} and reversible shear jamming (SJ) into a state with finite yield stress \cite{Scott2012,Ness2015,Ivo-Nature2016,Han2016,Sayantan2017}. Recent studies identified a stress-activated crossover from hydrodynamic interactions to frictional particle contacts to be key for these behaviors \cite{Seto2013,Fernandez2013,Cates-friction,Mari2015,Royer2015,Mari2014,Ness2015,Guy2015}. However, many suspensions exhibit only DST and not SJ. Here we show that particle surface chemistry can play a central role in creating conditions that allow for SJ. We find the system's ability to form interparticle hydrogen bonds when sheared into contact elicits SJ. We demonstrate this with charge-stabilized polymer microspheres and non-spherical cornstarch particles, controlling hydrogen bond formation with solvents. The propensity for SJ is quantified by tensile tests \cite{Sayantan2017} and linked to an enhanced friction by atomic force microscopy.  Our results extend the fundamental understanding of the SJ mechanism and open new avenues for designing strongly non-Newtonian fluids.
\vspace{2mm}
\newline
\hrule
\vspace{5mm}
In suspensions, increasing the particle packing volume fraction $\phi$ will increase the viscosity $\eta$ until, at some critical $\phi_J$, the fluid jams and $\eta$ diverges \cite{Barnes1989,Brown2014Rev,Mari2015,Wagner2013}. However, this is not the only route to solidification.  Suspensions prepared in an unjammed state can also solidify under an applied shear stress. At fixed packing fraction $\phi < \phi_J$, increasing the applied shear stress drives the suspension from a lubrication-dominated regime, through a DST regime, into a  shear-jammed state with non-zero yield stress. This behavior was predicted for frictional particles \cite{Cates-friction, Behringer2011} and recently observed in experiments on dense suspensions \cite{Ivo-Nature2016,Han2016,Sayantan2017}.  Cornstarch in water is a prototypical suspension to exhibit such a transition from DST to SJ. However, while dense suspensions of hard, non-aggregating particles can generically exhibit pronounced DST at high packing fractions \cite{Brown2010}, they do not necessarily shear jam.  Which particle characteristics enable SJ of suspensions has so far been unresolved.

\begin{figure}[h!]
	\centering
	\includegraphics[width=0.7\linewidth]{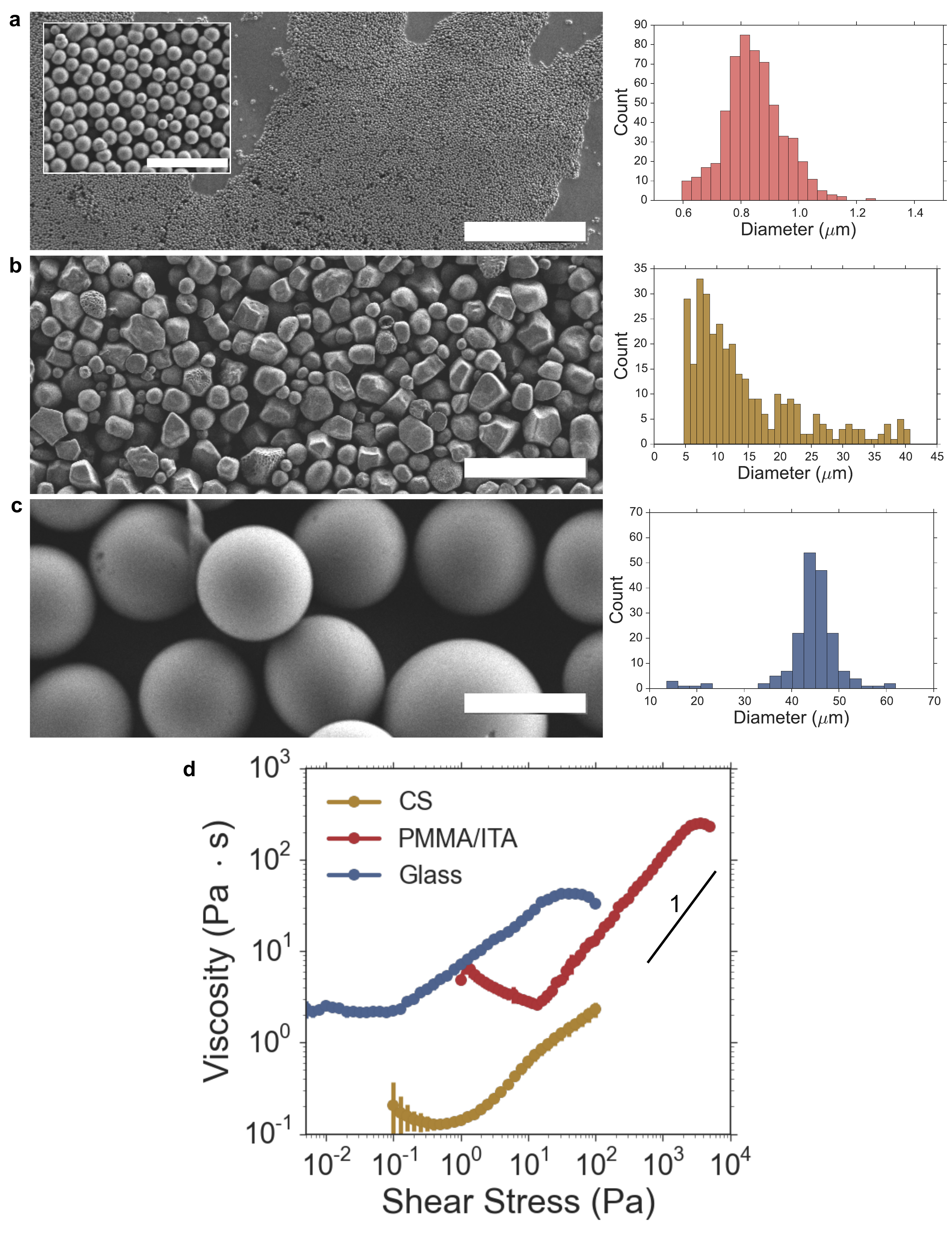}
	\caption{\textbf{Particle characteristics and flow curves}.  Histograms of particle size distribution obtained from image analysis of three particle systems: (a) PMMA/ITA, (b) cornstarch (CS), (c) glass beads. Images were taken by scanning electron microscopy. Scale bars indicate 50 $\mu$m, except in the inset to (a), where the scale bar denotes 5 $\mu$m. (d) Viscosity as a function of applied shear stress measured for PMMA/ITA suspended at volume fraction $\phi$=56\% in 69\% aqueous glycerol (aq. gly.), cornstarch suspended at $\phi$=43\% in 50\% aq. gly., and glass beads suspended at $\phi$=60\% in 100cSt silicone oil. Error bars represent the standard deviation from 3 replicate measurements.}
	\label{fgr:Figure1}

\end{figure}

To address this, we examine three types of particles in suspension. Comparing cornstarch to poly(methyl-methacrylate)/itaconic acid (PMMA/ITA) microspheres and glass beads allows us to test the roles of size dispersion, shape, and material. Figure \ref{fgr:Figure1} compares the particle morphology and steady-state suspension rheology. All three systems exhibit strong, nearly discontinuous shear thickening (a slope of unity in a log-log plot of viscosity versus shear stress corresponds to a discontinuous jump in viscosity when plotted as a function of shear rate). Only the cornstarch and PMMA/ITA suspensions, however, exhibit SJ. 

\begin{figure}[t]
	\centering
	\includegraphics[width=\linewidth]{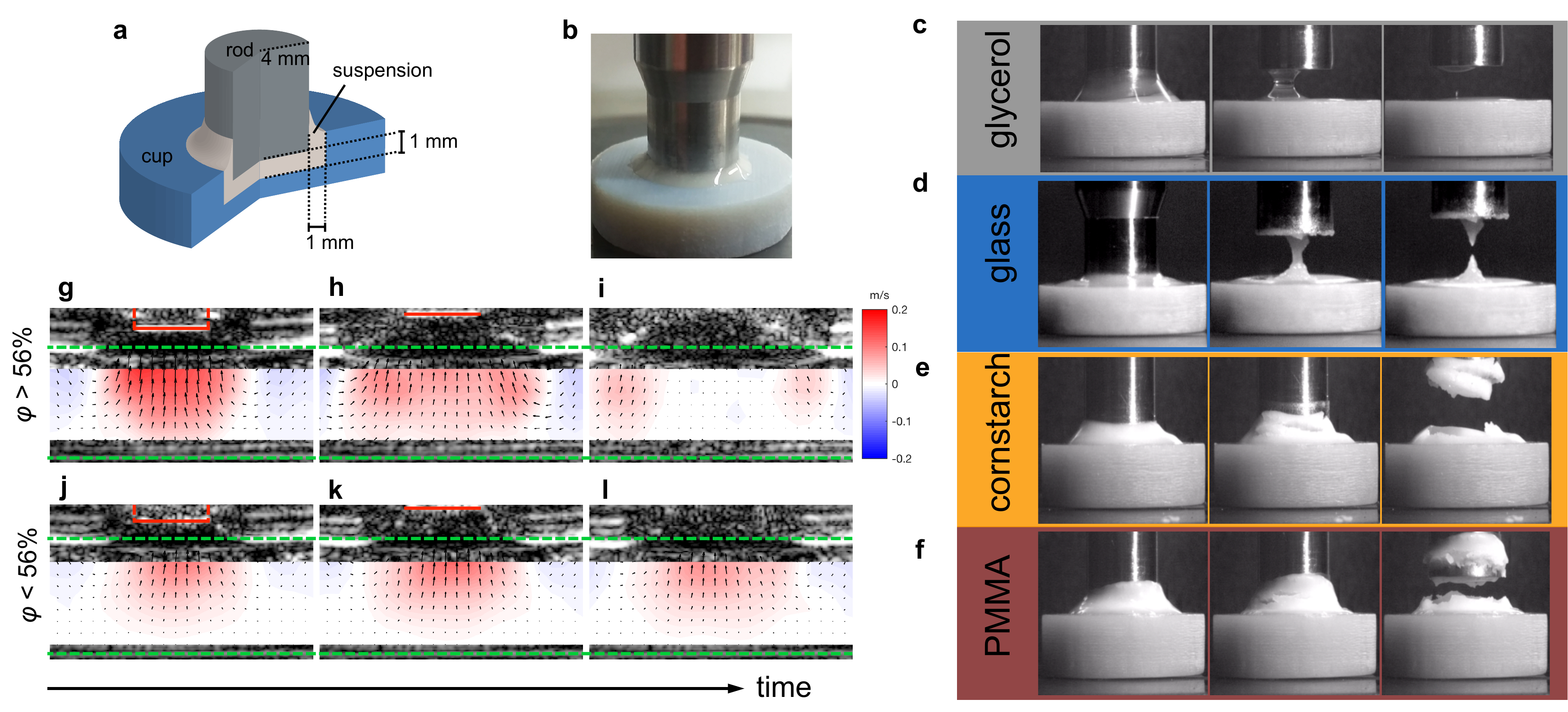}
	\caption{\textbf{Tensile testing for shear jamming}. (a) Schematic of extension test cell, shown in cross-section. The rod is connected to a rheometer. (b) Image of extension-test cell, with loaded sample. (c-f) Images of the extensional behavior of glycerol, glass, cornstarch, and PMMA, respectively. In each row, the suspension is depicted prior to the start of the measurement, near the point of peak force and incipient failure, and and after rupture of the fluid column. Full videos can be found in the Supplementary Information. (g-i) High-speed ultrasound images of propagating shear jamming fronts in concentrated ($\phi >$ 56\%) PMMA/ITA suspension under tensile stress. Velocity maps from particle image velocimetry are overlaid. Color field indicates vertical velocity, with red indicating upward motion, and blue indicating downward motion. The jamming front corresponds to the perimeter of the jammed (red) region\cite{Han2016}. The top and bottom boundaries of the suspension are indicated by dashed green lines. The rod pulling the suspension upward is indicated by the red line. The images are snapshots in time, showing the jamming force reaching (g) and interacting with the bottom boundary (h) until a non-moving, solid-like state is achieved (i), indicated by the large white, zero-velocity region. (j-l) High-speed ultrasound images of PMMA/ITA suspension which does not exhibit shear jamming ($\phi <$ 56\%). This time sequence shows only viscous flow.}
	\label{fgr:Figure2qual}
\end{figure}

As a method to quickly identify SJ when scanning across a wide range of suspension parameters, we take advantage of the finding that a shear-jammed state can support significant tensile stresses, and fracture as a solid \cite{Sayantan2017,Smith2010}. This is shown qualitatively in Fig. \ref{fgr:Figure2qual}, which compares results from tensile tests (Fig \ref{fgr:Figure2qual}a-b). Visual inspection of the suspension column as it is pulled apart shows that non-SJ suspensions (Fig. \ref{fgr:Figure2qual}c-d) experience a neck and pinch-off detachment. However, SJ suspensions (Fig. \ref{fgr:Figure2qual}e-f) show cleavage planes suggesting brittle fracture. This approach is similar to the extensional rheology studies done by Bischoff-White \emph{et al}. \cite{BischoffWhite2009}, however we measure normal force instead of viscosity. This allows us to differentiate solidification from shear thickening.  

A direct demonstration that this solidification is due to stress-activated shear jamming is achieved by observing the associated, rapidly-propagating jamming fronts. Such fronts have previously been identified as driving suspensions into the SJ state under impact \cite{Scott2012,Peters2D,Ivo-Nature2016}, and during the initial phase of applying Couette shear \cite{Ivo-Nature2016}. Importantly, they have also been identified as generating the jammed solid that emerges during the application of tensile stress in cornstarch suspensions \cite{Sayantan2017}. Using high-speed ultrasound imaging we find the same jamming fronts in the PMMA/ITA system as well (Fig. \ref{fgr:Figure2qual}g-i). Note that, while the suspension surface is pulled upward, these fronts propagate downward into the bulk of the suspension, transforming it into solid-like block of material that ruptures at large strain (Fig. \ref{fgr:Figure2qual}f). Stress is supplied by pulling on the suspension, with fixed speed 8mm/s in our experiments, and balanced by viscous stress at the the position of the moving front, where shear is localized \cite{Ivo-Nature2016,Han2016,Sayantan2017,Peters2D}. To enter the SJ regime, both a minimum stress level (or, equivalently, a minimum extension rate \cite{Smith2010}) and a sufficiently high packing fraction are required \cite{Ivo-Nature2016}. For the PMMA/ITA system this threshold lies at $\phi \simeq$ 56\%. Less concentrated suspensions exhibit viscous flow but no propagating fronts (Fig. \ref{fgr:Figure2qual}j-l).

\begin{figure}[h]
	\centering
	\includegraphics[width=0.5\linewidth]{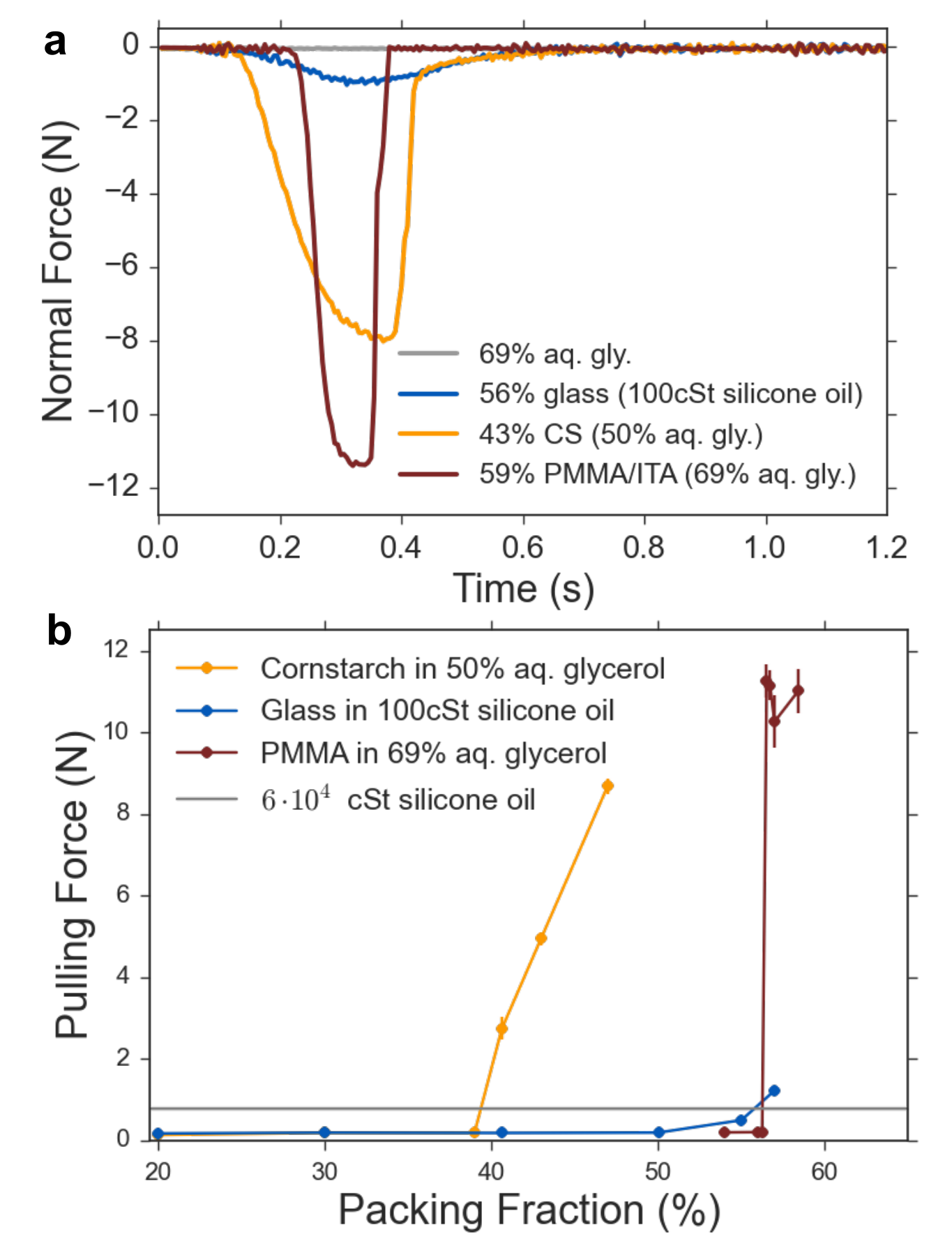}
	\caption{\textbf{Quantative analysis of shear jamming tensile tests}. (a) Representative raw data for a variety of suspension systems, as indicated, showing the normal force measured by the rheometer as the rod is pulled upward at a rate of 8 mm/s. (b) Magnitude of peak force from data in (a), as a function of volume fraction. High-viscosity silicone oil is shown for comparison. Error bars represent the standard deviation from 5 replicate measurements.}
	\label{fgr:Figure2quant}
	
\end{figure}

Figure \ref{fgr:Figure2quant} shows more quantitatively the response to tensile strain as particle type and packing fraction are varied. Suspensions that enter the SJ regime produce strikingly different forces during extension than non-SJ suspensions (Fig. \ref{fgr:Figure2quant}a). To characterize these differences, we plot in Fig. \ref{fgr:Figure2quant}b the pulling force, defined here as the peak magnitude of the normal force, as a function of packing fraction. Even at 57\%, the glass bead suspensions display forces commensurate with that of oils near the same effective viscosity (6$\cdot$10$^{6}$cSt silicone oil). On the other hand, both cornstarch and PMMA/ITA reach pulling forces of  8N or larger, corresponding to stresses in excess of 100kPa. These stress levels are significantly larger than achievable under steady-state flow (Fig. \ref{fgr:Figure1}d) and are a quantitative indication of solidification \cite{Sayantan2017}.

The results in Figs. \ref{fgr:Figure1} and \ref{fgr:Figure2qual} rule out particle shape and size as primary drivers for SJ: smooth, size-controlled PMMA/ITA microspheres around 1 $\mu$m in diameter exhibit SJ just as well and even more strongly than rough, irregular cornstarch particles with a median diameter near 11 $\mu$m. This apparent independence on particle geometry drives us to examine more closely the roles of particle surface chemistry and solvent.

Starch particles have surface hydroxyl groups, while PMMA/ITA particles have surface carboxyl groups.  In aqueous suspensions, this allows both starch and PMMA/ITA  to form interparticle hydrogen bonds. Here we show that this chemical contribution to the nature of interparticle contacts can elicit SJ. To test this, we investigate the effect of adding a chaotropic agent, \emph{i.e.} a compound that specifically interferes with hydrogen bonding,such as urea, to the suspending solvent. 

\begin{figure}[h!]
	\centering
	\includegraphics[width=0.45\linewidth]{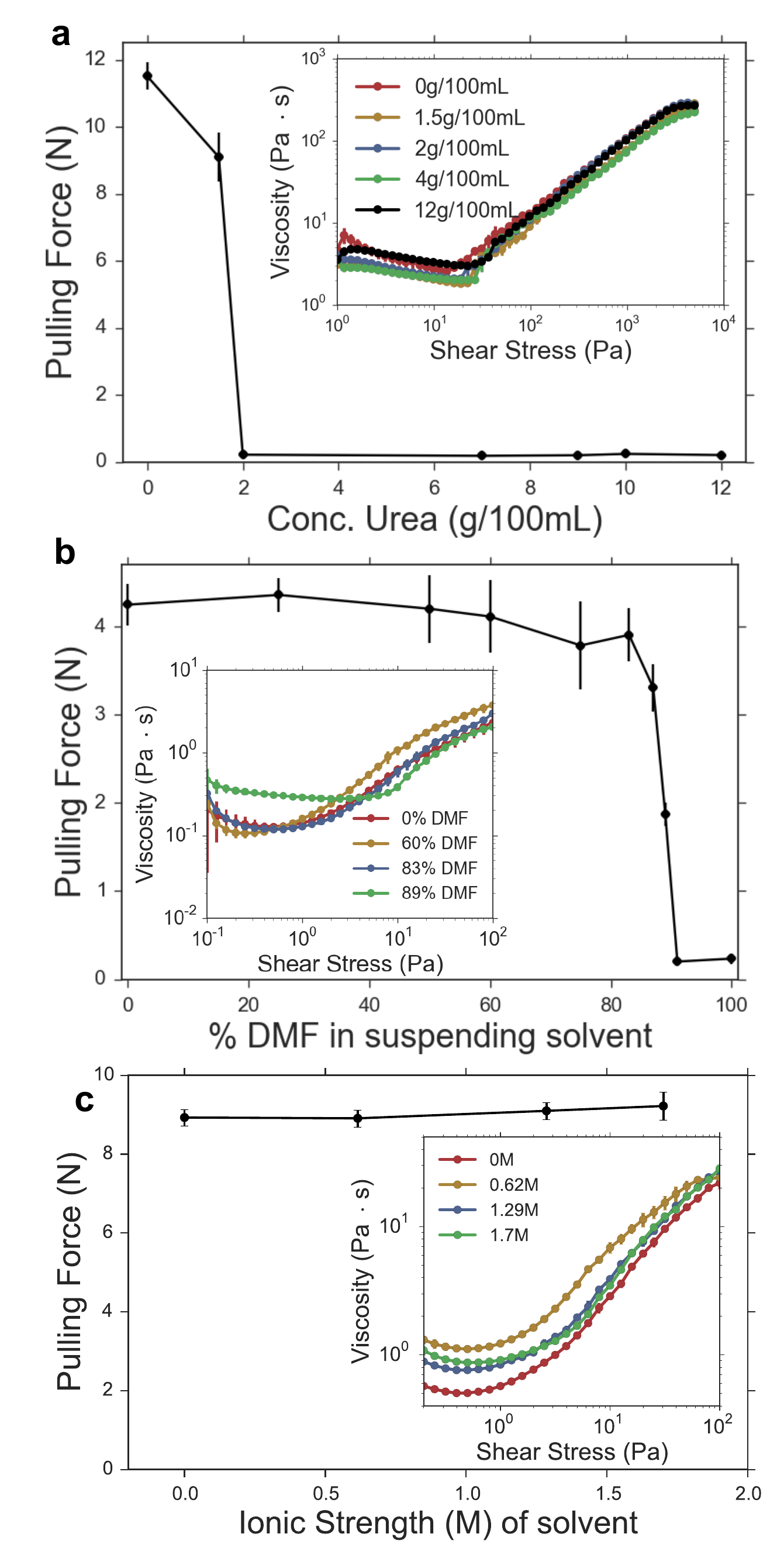}
	\caption{\textbf{Shear jamming dependence on hydrogen bonding}. (a) Suppressing shear jamming by adding urea, a chaotropic agent that interferes with hydrogen bonds. Data show the peak pulling force as a function of urea concentration for $\phi$ = 57\% PMMA/ITA particles in in 69\% aq. gly. Error bars represent the standard deviation from 5 replicate measurements. Inset: steady-state rheometry data for the tested suspensions. Error bars represent the standard deviation from 3 replicate measurements. (b) Suppressing shear jamming by adding DMF, a polar aprotic solvent.  Data show the peak pulling force as a function of DMF concentration for $\phi$ = 43\% cornstarch in 50\% aq. gly. Error bars represent the standard deviation from 5 replicate measurements.  Inset: steady-state flow curves for the tested suspensions. Error bars represent the standard deviation from 2-3 replicate measurements. (c) Cornstarch/water suspension SJ is independent of ionic strength. Data show the peak pulling force of $\phi$ = 43\% cornstarch suspensions in 50\% aq. gly., as a function of ionic strength from adjusting the NaCl concentration. }
	\label{fgr:Figure3}
\end{figure}

Figure \ref{fgr:Figure3}a shows the pulling force as a function of urea concentration for PMMA/ITA suspensions. At our applied shear rate, shear jamming is eliminated above concentrations of $\sim 2$  g/100mL in a  $\phi=56\%$ suspension. This threshold concentration corresponds to an order of $10^{22}$ urea molecules, commensurate with the number of hydrogen bonding sites in the system: the particle surface density of COOH groups is around $10$nm$^{-1}$ \cite{Appel2013}, which implies for the $0.845$ $\mu$m diameter particles $\sim 10^{22}$ surface COOH groups per 100ml of $\phi=56\%$ suspension.   The inset in Fig. \ref{fgr:Figure3} a shows that the steady-state suspension rheology, and thus DST, remains unchanged across all urea concentrations.

Fig. \ref{fgr:Figure3}b shows the pulling response of $\phi$ = 43\% cornstarch in water/dimethylformamide (DMF) mixtures. Water is a polar protic solvent, capable of donating hydrogen bonds. DMF is a polar aprotic solvent, which has no polar hydrogens and thus is not a hydrogen bond donor. DMF has a viscosity and density very close to that of water (0.92 mPa$\cdot$s and 0.944 g/mL at 20\degree C, respectively). By tuning the concentration of water in water/DMF mixtures, we tune the capacity of the suspending solvent to mediate interparticle hydrogen bonds as the particles come into contact under applied shear, or separate after the shear is released. As the DMF concentration is increased, no change in the pulling force is seen up until $\sim$ 80\% DMF, after which the pulling force abruptly drops below the detection threshold. We interpret this to indicate that as long as sufficient water is available to mediate interparticle hydrogen bonding, shear jamming remains possible. If the polar protic solvent activity drops too low, hydrogen bonding interactions are not fully mediated and shear jamming becomes suppressed.

Adding a salt to cornstarch suspensions does not change the tensile strength (Fig. \ref{fgr:Figure3}c), even at ionic strengths of 1.7M, where the Debye length is reduced to the scale of single molecules. To be unaffected by salt concentration, the chemical interactions must take place at length scales smaller than the Debye length. This is an indicator that the hydrogen bonding occurs at locations where particles come into extremely close contact.

\begin{figure}[h!]
	\centering
	\includegraphics[width=0.4\linewidth]{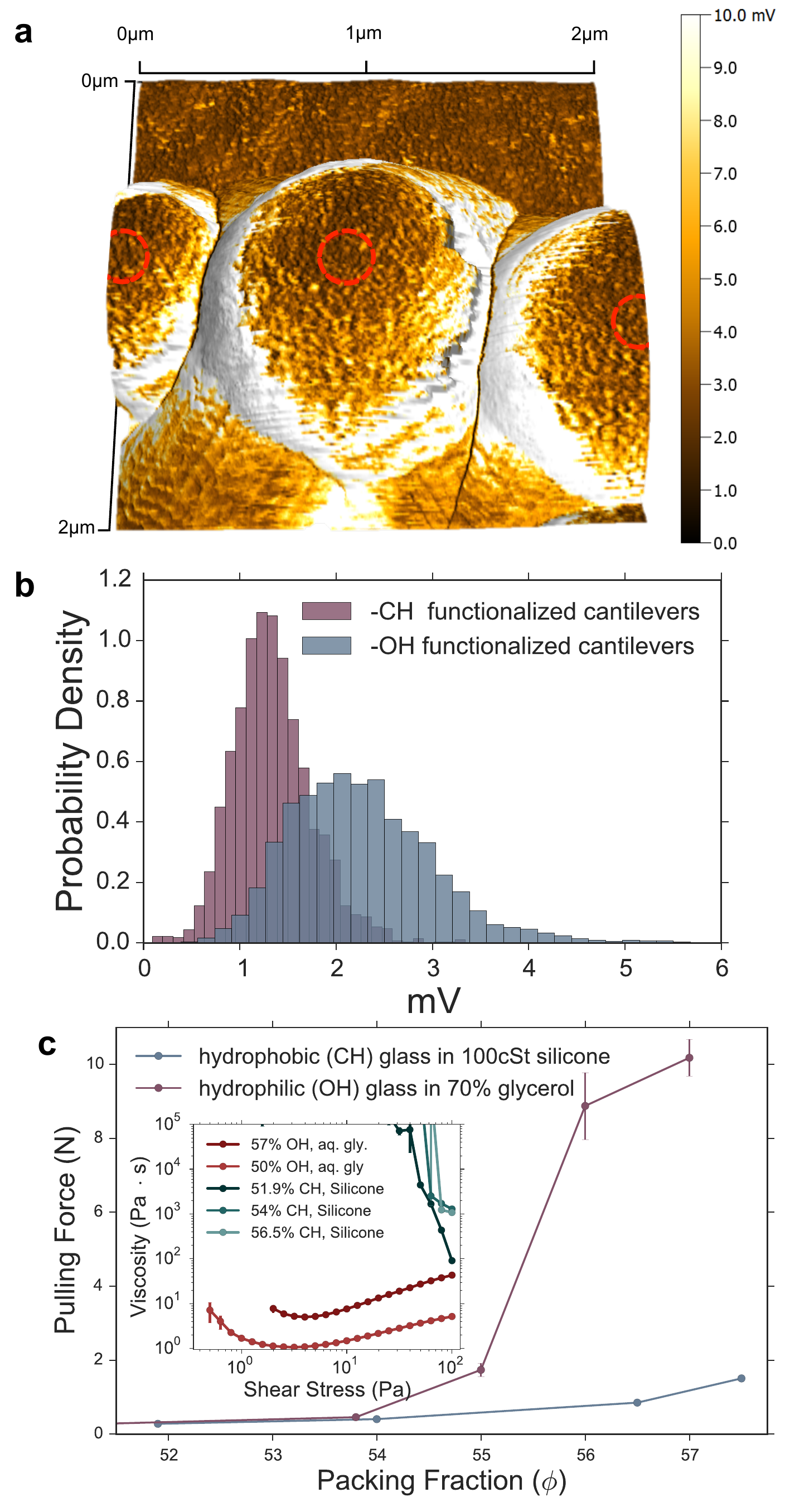}
	\caption{\textbf{Connecting particle surface chemistry and interparticle friction}. (a) AFM height image of PMMA/ITA particles overlaid with a color map that indicates the lateral force response, here shown for the case of an OH-terminated AFM cantilever. The raw lateral force data (in units of mV) is directly related to the magnitude of the friction coefficient. Red dashed circles indicate the areas over which the data were analyzed for comparison with data obtained using a CH-terminated cantilever. These regions  around the particles' apex were selected to avoid geometric effects and tip shape artifacts as the AFM tip moves near a particle edge. (b) Histograms of extracted lateral force magnitudes, comparing CH-terminated (purple) and OH-terminated (blue) AFM cantilevers. OH-terminated cantilevers exhibit a mean force of 2.26  $\pm$0.74 mV, which drops to 1.33mV $\pm$0.41 mV for CH-terminated cantilevers. Surface functionalization of the gold-coated AFM cantilevers was achieved by ethanethiol treatment (for CH termination) or mercaptoethanol treament (for OH termination), as described in the Methods. (c) Pulling force as a function of packing fraction for 3-6um glass beads, with different surface treatments. While hydrophobic, silanized soda-lime glass beads in 100cSt silicone oil (blue) show no stronger force than a highly-viscous fluid, shear jamming is observed in hydrophilic, silanol-terminated glass beads in aq. gly. (red). Error bars represent the standard deviation from 5 replicate measurements. Inset: steady-state rheology of suspensions. Error bars represent the standard deviation from 2 replicate measurements.}
	\label{fgr:FigureAFM}
\end{figure}

Hydrogen bonding provides a short-range attractive force on the scale of $\sim$ 5 kJ/mol, similar to a weak chemical bond, while still being reversible under thermal fluctuations in polar protic solvents like water. The associated small normal forces have recently been measured directly by atomic force microscopy (AFM) \cite{Galvez2017}. For enabling shear jamming, prior theoretical work \cite{Cates-friction,Ness2015} has suggested that particularly strong frictional (tangential) forces are required. We therefore explored the repercussions of hydrogen bonding on friction by operating an AFM in lateral force mode with cantilevers that were chemically surface-functionalized with either hydroxyl (OH) or alkyl (CH) groups. This strategy allowed us to vary the surface chemistry of only one surface, isolating its effect. Figure \ref{fgr:FigureAFM}a shows the height image of a group of PMMA/ITA particles glued to a substrate, with friction magnitude represented by color. From analysis of the two cases we find (Fig. \ref{fgr:FigureAFM}b) that when the cantilever was OH-terminated to enable hydrogen bonding, the result is a nearly two-fold enhancement of the average strength of friction. 

From these results a picture emerges in which shear jamming is activated when particles are forced sufficiently close that interparticle hydrogen bonds can create the equivalent of strong frictional interactions. The fact that we can deactivate SJ (at our extension rate of 8mm/s) via the solvent while leaving DST essentially unaffected (Fig. \ref{fgr:Figure3}a,b) highlights the difference between the two phenomena. Our findings are in line with recent models where SJ emerges only once a sufficiently large fraction of particles have crossed over from interacting via frictionless hydrodynamic lubrication to undergoing enduring frictional contacts so that the suspensions solidifies and exhibits a yield stress \cite{Cates-friction,Ness2015}. By contrast, DST in dense suspensions can be observed after particles break through the lubrication layer as long as particles exhibit just enough friction that steady-state shear will generate enhanced stress due to "frustrated dilation" \cite{Brown2014Rev,Brown2010} or, equivalently, intermittent, partial jamming events \cite{Seto2013,Fernandez2013,Mari2015,Mari2014}.

In prior work on aqueous suspensions of fine silica particles, hydroxyl groups at the particle surfaces have been associated with increased shear thickening through hydrogen bonding with the suspending solvent \cite{Qin2016RSC,Warren2015}. However, the details of how the resulting solvation layer would create shear-induced viscosity increases have remained unclear. Here the notion of inter-particle hydrogen bonds providing strong frictional interactions brings a new perspective. Previously, the mild thickening behavior of dilute polysaccharide solutions (not suspensions) was linked to shear-induced creation of inter-molecular hydrogen bonds\cite{Jaishankar2015}. On the scale of molecules, the nature of these interactions manifest as cohesion, while for much larger particles this interaction may manifest primarily as enhanced friction. Indeed, models and simulations highlight the key role of friction, finding SJ without invoking cohesion \cite{Cates-friction,Ness2015}; although, for frictionless systems weak attractive interactions seem required \cite{Ning}.  

We can take these results one step further and use them to rationally design systems that highlight or eliminate SJ behavior.  For example, hydrophilic soda lime glass has surface silanol (Si-OH) groups, and thus can form interparticle hydrogen bonds. Therefore, suspensions of these particles in water/glycerol mixtures can be expected to shear jam. However, hydrophobic, silane-functionalized glass spheres of the same size are incapable of interparticle hydrogen bonding. When dispersed in an appropriate sovlent (\emph{e.g.} silicone oil), the suspensions may display non-Newtonian behavior, but they are not expected to SJ. To test this, we used commercially-available 3-6$\mu$ glass spheres (Cospheric). The results are shown in Fig. \ref{fgr:FigureAFM}c, and confirm these predictions. Thus, by carefully considering the particle surface chemistry and solvent, we can design shear jamming capacity. 

A key aspect emerging from the data presented here is that, while both DST and SJ rely on direct particle-particle contacts, the stress-activated solidification associated with SJ requires an additional step that qualitatively enhances the contact interaction. We have shown that this can be provided by hydrogen bonding between particles. Geometric variables such as particle shape and size appear not to pose fundamental requirements for SJ, although they may affect details of the behavior. 

This explains why suspensions of starch particles exhibit DST as well as SJ, while other particle systems only show DST unless their surfaces are suitably conditioned.  It also explains why we can suppress SJ by selectively preventing hydrogen bonding.  Our work highlighted cornstarch particles, because of their frequent use in experiments on the non-Newtonian behavior of dense suspensions \cite{Fall2015,Boyer2016,Jerome2016}. However, as the data on PMMA/ITA and glass spheres demonstrate, the ability to control SJ through the particle surface chemistry is more general. This provides a powerful new tool for the design of suspension-based materials that can transform reversibly from fluid to solid-like in response to applied stresses.

\textbf{Acknowledgements}
We thank S. Majumdar, I. Peters, and M. Lueckeheide for insightful discussions. This work was
supported by the  University of Chicago Womens Board, and the US Army Research Office through grant W911NF-16-1-0078. Additional support was provided by the Chicago MRSEC, which is funded by the NSF through grant DMR-1420709.

\section*{\centering{Methods}}

\textbf{PMMA/ITA particle synthesis}
We synthesized micron-scale poly(methylmethacrylate)/itaconic
acid (PMMA/ITA) particles according to the procedure reported by Appel \emph{et al.} \cite{Appel2013}. Briefly: methyl methacrylate (MMA), water, and itaconic acid are combined in a round-bottom flask. The flask is sealed with a septum and purged with nitrogen. After heating to 85\degree C while stirring, 4,4’-Azobis(4-cyanovaleric acid) is injected to initiate the reaction. The reaction is allowed to proceed for 12 hours, after which the particles can be separated by centrifugation and dried. The precise particle size is controlled by the MMA/water ratio (55g/125g for 1$\mu$m diameter particles). Particles were imaged by scanning electron microscopy (SEM) using a Zeiss Merlin SEM. Particle surface charge was obtained by dynamic light scattering (DLS) using a Malvern ZetaSizer Nano ZS. Particle sizes were obtained by DLS and SEM image analysis.

\textbf{Suspension preparation}
PMMA and glass suspensions are reported as volume fractions: suspending solvent and dry particles were measured out by mass, and converted to volume through density. To obtain precise packing fractions, suspending solvent was weighed first, and particles were added directly to solvent while recording mass. Suspensions were allowed to equilibrate for at least 45 minutes before use, and were used same-day to avoid solvent evaporation. 

Cornstarch particles have a porosity around 30\% and thus absorb solvent, which significantly raises the actual volume fraction of solids. Han \emph{et al}. carefully characterized the solvent-dependent absorption to obtain corrected volume fractions \cite{Han2017-inpress}. Due to the variety of solvents used in the experiments reported here, we do not perform these corrections and  for cornstarch simply list the uncorrected solids fraction based on the dry material volume dispersed. 

\textbf{Tensile test protocol}
We used an Anton Paar MCR301 rheometer to pull on the suspension surface by withdrawing an 8 mm rod that was partially embedded to a depth of 1 mm in the sample, while measuring the normal force response on the rod. In these experiments, the rheometer only applied a normal force; there was no rotation of the rod. A schematic and image of the apparatus with dimensions is shown in Fig. \ref{fgr:Figure2qual} a-b. To ensure the sample volume and contact with the rod was consistent between experiments, a 3-D printed cup was used to contain the sample. The cup was affixed to the bottom plate of the rheometer with double-sided tape, and the suspension was leveled to the top of the cup prior to lowering the rod. All tests were run at a constant vertical extension rate of 8 mm/s. 

\textbf{Steady state rheology}
Steady-state rheological experiments were conducted using a stress-controlled Anton Paar MCR301 rheometer with parallel plate (25 mm diameter) geometry. Sample temperatures were
maintained to within 1\degree C for each run. All experiments were conducted within 22-25 \degree C. A solvent trap was used to mitigate sample evaporation.

\textbf{Ultrasound Imaging} 
The bulk suspension flow field under extension was imaged using high-speed ultrasound following the method reported by Han \emph{et al.} \cite{Han2016}. The PMMA/ITA suspension ($\phi \simeq$ 57\%) was degassed and loaded into an impedence-matched, cylindrical container (d = 5 cm) to a depth of 1.7 cm. A cylindrical rod (d = 0.6 cm), controlled by a linear actuator, was used to pull the suspension at a velocity of 200 mm/s. The suspension flow was visualized from below, using the reflection of ultrasound off of suspension bubbles as tracer particles. Particle imaging velocimetry of these videos was used to obtain the flow field.

\textbf{AFM procedure} 
An Asylum Research MFP-3D-BIO atomic force microscope operating in lateral force mode was used to acquire frictional force images. Gold-coated silicon-nitride AFM cantilevers (Olympus RC800PB, type 4: k = 0.11 N/m; tip radius = 30nm) were calibrated for spring constant and sensitivity. Cantilever tips were chemically functionalized by washing in ethanol, isopropanol, and acetone, followed by dipping in either ethanethiol (CH-functionalized), or mercaptoethanol (OH-functionalized). Particles were immobilized in a thin layer of epoxy (Hardman no. 04001) on top of a glass coverslide, and left to dry overnight. Images were acquired in water at a 1 Hz line scan rate and 0.9 nN load.


\end{document}